\documentclass[10pt]{iopart}

\usepackage{amsfonts,amsthm}

\usepackage{times,mathptmx}

\theoremstyle{plain}

\newtheorem*{prop}{Proposition}
\newtheorem*{lemma}{Lemma}

\theoremstyle{remark}

\newtheorem*{rem}{Remark}

\theoremstyle{definition}

\newtheorem*{ex}{Example}

\newcommand{\CC}{\mathbb{C}}

\newcommand{\RR}{\mathbb{R}}

\newcommand{\HH}{\mathcal{H}}

\newcommand{\LL}{\mathcal{L}}

\newcommand{\NN}{\mathcal{N}}

\newcommand{\ran}{\mathop{\mathrm{ran}}}

\newcommand{\dom}{\mathop{\mathrm{dom}}}

\newcommand{\res}{\mathop{\mathrm{res}}}

\newcommand{\gr}{\mathop{\mathrm{gr}}}

\newcommand{\spec}{\mathop{\mathrm{spec}}}

\newcommand{\diag}{\mathop{\mathrm{diag}}}

\newcommand{\eqref}[1]{\textup{(\ref{#1})}}

\newcommand{\text}[1]{\mbox{#1}}

\renewcommand{\Im}{\mathop{\mathrm{Im}}}

\renewcommand{\Re}{\mathop{\mathrm{Re}}}

\begin{document}

\title[]{A remark on Krein's resolvent formula and boundary conditions}

\author{Sergio Albeverio\dag{} and Konstantin Pankrashkin\ddag}

\address{\dag\ Institut f\"ur Angewandte Mathematik, Rheinische
Friedrich-Wilhelms-Universit\"at Bonn, Wegelerstrasse~6, 53115 Bonn, Germany;
BiBoS Research Center, Bielefeld, Germany; CERFIM Locarno;
Accademia di Architettura, Universit\`a della Svizzera Italiana, Mendrisio,
Switzerland; Dipartimento di Matematica, Universit\`a di Trento, Trento, Italy
\\
\ddag\ Institut f\"ur Mathematik, Humboldt-Universit\"at zu Berlin,
Unter den Linden~6, 10099 Berlin, Germany (corresponding author);
E-mail: \mailto{const@mathematik.hu-berlin.de}}

\begin{abstract}
We prove an analog of Krein's resolvent formula expressing the
resolvents of self-adjoint extensions in terms of boundary
conditions. Applications to quantum graphs and systems with point interactions are discussed.
\end{abstract}





\nosections

Krein's resolvent formula~\cite{krein} is a powerful tool in the
spectral analysis of self-adjoint extensions, which found numerous
applications in many areas of mathematics and physics, including
the study of exactly solvable models in quantum
physics~\cite{alb,pavlov,ak}. For the use of this formula in the
traditional way one needs a kind of preliminary construction, like
finding a maximal common part of two extensions,
see~\cite[Appendix~A]{alb}. While this is enough for many
applications, including models with point interactions, there
is a number of problems like the study of quantum graphs or
more general hybrid structures, where self-adjoint extensions are
suitable described by more complicated boundary conditions,
see~\cite{ks,bg,exner}, and it is necessary to modify Krein's
resolvent formula to take into account these new needs. This can
be done if one either modifies the coordinates in which the
boundary data are calculated~\cite{bg,harmer} or considers
boundary conditions given in a non-operator way using linear
relations~\cite{gorb,dm}. On the other hand, a more
attractive idea is to have a resolvent formula taking directly the
boundary conditions into account, without changing the
coordinates. We describe the realization of this idea in the
present note.

Let $S$ be a closed densely defined symmetric operator with the
deficiency indices $(n,n)$, $0<n<\infty$, acting in a certain
Hilbert space $\HH$. One says that a triple
$(V,\Gamma_1,\Gamma_2)$, where $V=\CC^n$ and $\Gamma_1$ and
$\Gamma_2$ are linear maps from the domain $\dom S^*$ of the adjoint of $S$
to $V$, is a
\emph{boundary value space} for $S$ if $\langle \phi,
S^*\psi\rangle- \langle S^*\phi, \psi\rangle= \langle \Gamma_1
\phi,\Gamma_2\psi\rangle-\langle \Gamma_2
\phi,\Gamma_2\psi\rangle$ for any $\phi,\psi\in\dom S^*$ and the
map $(\Gamma_1,\Gamma_2):\dom S^*\to V\oplus V$ is surjective. A
boundary value space always exists~\cite[Theorem 3.1.5]{gorb}. All self-adjoint
extensions of $S$ are restrictions of $S^*$ to functions
$\phi\in\dom S^*$ satisfying $A\Gamma_1\phi=B\Gamma_2\phi$, where
the matrices $A$ and $B$ must obey the following two properties:
\begin{eqnarray}
                 \label{eq-ks1}
AB^*=BA^* \quad\Leftrightarrow\quad A B^* \text{ is self-adjoint},\\
                 \label{eq-ks2}
\text{the $n\times 2n$ matrix $(A|B)$ has maximal rank $n$}.
\end{eqnarray}
We denote such an extension of $S$ by $H^{A,B}$. Our aim is to
write a formula for the resolvent $R^{A,B}(z)=(H^{A,B}-z)^{-1}$ in
terms of these two matrices $A$ and $B$.

We will need some notions from the theory of linear relations. Let
$V=\CC^n$. Any linear subspace $\Lambda$ of $V\oplus V$ will be
called a \emph{linear relation} on $V$. By the \emph{domain} of
$\Lambda$ we mean the set $\dom\Lambda=\{x\in V:\,\exists y\in V
\text{ with } (x,y)\in\Lambda)\}$. A linear relation
$\Lambda^{-1}=\{ (x,y):\,(y,x)\in\Lambda\}$ is called \emph{inverse}
to $\Lambda$. For $\alpha\in\CC$ we put $\alpha \Lambda=\{(x,\alpha
y):\,(x,y)\in\Lambda\}$.
 For two linear relations
$\Lambda',\Lambda''\subset V\oplus V$ one can define the sum
$\Lambda'+\Lambda''=\{(x,y'+y''),\,(x,y')\in\Lambda',\,
(x,y'')\in\Lambda''\}$; clearly, one has $\dom
(\Lambda'+\Lambda'')=\dom\Lambda'\cap\dom\Lambda''$. The graph of
any linear operator $L$ acting in $V$ is a linear relation, which
we denote by $\gr L$. Clearly, if $L$ is an invertible operator,
then $\gr L^{-1}=(\gr L)^{-1}$. For arbitrary linear operators
$L',L''$ one has $\gr L'+\gr L''=\gr (L'+L'')$. Therefore, the set
of linear operators is naturally imbedded into the set of linear
relations.

Denote by $J$ an operator acting in $V\oplus V$ by the rule
$J(x_1,x_2)=(x_2,-x_1)$, $x_1,x_2|in V$. For a linear relation $\Lambda$ on $V$
the relation $\Lambda^*=J\Lambda^\bot$ is called \emph{adjoint} to
$\Lambda$; $\Lambda$ is called \emph{symmetric} if $\Lambda\subset
\Lambda^*$ and is called \emph{self-adjoint} if
$\Lambda=\Lambda^*$. The graph of a linear operator $L$ in $V$ is
symmetric (respectively, self-adjoint), iff its graph is a
symmetric (respectively, self-adjoint) linear relation. In other
words, a self-adjoint linear relation (abbreviated as s.a.l.r) is
a symmetric linear relation of dimension $n$. Let $A$, $B$ be
$n\times n$ matrices. We introduce the notation
\[
\Lambda^{A,B}=\big\{ (x_1,x_2)\in V\oplus V, \quad  A x_1=B x_2
\big\}.
\]
A criterion for $\Lambda^{A,B}$ to be self-adjoint was proven
in~\cite{ks}: \emph{A linear relation $\Lambda^{A,B}$ is
self-adjoint iff $A$ and $B$ satisfy \eqref{eq-ks1} and
\eqref{eq-ks2}.} It is important to emphasize that \emph{any}
s.a.l.r. $\Lambda$ can be defined by this construction, more
precisely, there exists a unitary operator $U$ such that
$\Lambda=\Lambda^{i(1+U),1-U}$~\cite{gorb}. This shows that there
is a bijection between s.a.l.r.'s and unitary operators;
nevertheless, we consider parametrization by the two matrices of
coefficients as a more natural way to present s.a.l.r.

The language of linear relations is widely used in the theory of
self-adjoint extensions of symmetric
operators~\cite{pavlov,harmer,nov}. Let us return to the
symmetric operator $S$ and its boundary value space
$(V,\Gamma_1,\Gamma_2)$. It is a well known fact that \emph{there
is a bijection between all self-adjoint extensions of $S$ and
s.a.l.r's on $V$. A self-adjoint extension $H^\Lambda$
corresponding to a s.a.l.r. $\Lambda$ is a restriction of $S^*$ to
elements $\phi\in\dom S^*$ satisfying abstract boundary conditions
$(\Gamma_1\phi,\Gamma_2\phi)\in\Lambda$}~\cite[Theorem 3.1.6]{gorb}. To carry out
the spectral analysis of the operators $H^\Lambda$ it is useful to
know their resolvents, which are provided by the famous Krein's
formula~\cite{krein,dm}. To write this formula we need some
additional constructions. For $z\in\CC\setminus\RR$, let $\NN_z$
denote the corresponding deficiency subspace for $S$, i.e.
$\NN_z=\ker(S^*-z)$. The restriction of $\Gamma_1$ and $\Gamma_2$
onto $\NN_z$ are invertible linear maps from $\NN_z$ to $V$. Put
$\gamma_z=\big(\Gamma_1|_{\NN_z}\big)^{-1}$ and
$Q(z)=\Gamma_2\gamma_z$; these maps form holomorphic families from
$\CC\setminus\RR$ to the spaces $\LL(V,\HH)$ and $\LL(V,V)$ of
bounded operators from $V\to\HH$ and from $V$ to $V$ respectively.
Denote by $H^0$ a self-adjoint extension of $S$ given by the
boundary conditions $\Gamma_1\phi=0$. i.e. for $A=1$ and $B=0$,
then the maps $\gamma_z$ and $Q(z)$ have analytic continuations to
the resolvent set $\res H^0$, and for all $z,\zeta\in\res H^0$ one
has~\cite{dm}
\begin{equation}
        \label{q-fun}
Q(z)-Q^*(\zeta)=(z-\overline\zeta)\,\gamma^{\,*}_\zeta\,\gamma_z.
\end{equation}
The maps $\gamma_z$  and $Q(z)$ are called the $\Gamma$-field and
the $\mathcal{Q}$-function for the pair $(S,H^0)$, respectively.

For all $z\in\res H^0\cap\res H^\Lambda$ we consider the linear
relation $\gr Q(z)-\Lambda$. While this set is, generally
speaking, not the graph of an operator, the inverse linear
relation $\big(\gr Q(z)-\Lambda\big)^{-1}$ is the graph of a
certain linear operator $C^\Lambda(z)$, so that the resolvent
$R^\Lambda(z)=(H^\Lambda-z)^{-1}$ is expressed through the
resolvent $R^0(z)=(H^0-z)^{-1}$ by \emph{Krein's
formula}~\cite[Proposition~2]{dm}
\begin{equation}
     \label{krein}
R^\Lambda(z)=R^0(z)-\gamma_z\, C^\Lambda(z)\,
\gamma^{\,*}_{\overline z}.
\end{equation}
The calculation of $C^\Lambda(z)$ is a rather difficult technical
problem, as it involves ``generalized'' operations with linear
relations. Such difficulties do not arise if $\Lambda$ is a graph of
a certain linear operator $L$; the corresponding boundary conditions
can be presented by
\begin{equation}
      \label{gamma-l}
\Gamma_2\phi=L\,\Gamma_1\phi,
\end{equation}
and such extensions are called \emph{disjoint with respect to
$H^0$} because they satisfy the equality $\dom H^\Lambda\cap\dom
H^0=\dom S$~\cite{krein} (the operator $S$ is then called the
\emph{maximal part} of $H^0$ and $H^\Lambda$). Then the subspace
$\gr Q(z)-\Lambda$ is the graph of the invertible \emph{operator}
$Q(z)-L$, and $C^\Lambda(z)=\big(Q(z)-L\big)^{-1}$. Actually, for
a \emph{given} self-adjoint extension $H$ one can find a boundary
value space (which is, of course, not unique) such that $H$
corresponds to the boundary conditions~\eqref{gamma-l} with a
suitable~$L$~\cite{bg,harmer}. But finding such boundary value
space involves a lot of other problems, in particular, the
operators $Q(z)$ and $\gamma_z$ must be
changed~\cite{dm,harmer,ms}. On the other hand, any boundary
conditions can be represented with the help of two matrices by
\begin{equation}
     \label{A-B}
A\Gamma_1\phi=B\Gamma_2\phi \quad \Leftrightarrow \quad
(\Gamma_1\phi,\Gamma_2\phi)\in\Lambda^{A,B}
\end{equation}
with $A$ and $B$ satisfying~\eqref{eq-ks1} and~\eqref{eq-ks2}. Our
aim is to show that the resolvent formula admits a simple form in
terms of these two boundary matrices. Here is the main result of
our note.
\begin{prop}[Modified Krein's resolvent formula]
       \label{prop-main}
Let $H^{A,B}$ be the self-adjoint extension of $S$ corresponding
to the boundary conditions~\eqref{A-B} with $A$, $B$
satisfying~\eqref{eq-ks1} and~\eqref{eq-ks2}, and $z\in\res
H^0\cap\res H^{A,B}$, then
\begin{itemize}
\item[(a)] the matrices $Q(z)B^*-A^*$ and $B Q(z)-A$ are
non-degenerate, \item[(b)] the resolvent
$R^{A,B}(z)=(H^{A,B}-z)^{-1}$ is connected with $R^0(z)$ by
\begin{eqnarray}
     \label{eq-krein1}
R^{A,B}(z)=R^0(z)-\gamma_z
B^*\big(Q(z)B^*-A^*\big)^{-1}\gamma^{\,*}_{\overline z},\\
\text{or}  \nonumber\\
     \label{eq-krein2}
R^{A,B}(z)=R^0(z)-\gamma_z
\big(BQ(z)-A\big)^{-1}B\,\gamma^{\,*}_{\overline z}
\end{eqnarray}
\end{itemize}
\end{prop}
\begin{rem}
The formulas~\eqref{eq-krein1} and~\eqref{eq-krein2} are
equivalent. To obtain them from each other one should replace $z$
by $\overline z$ and take in both sides adjoint operators taking
into account the resolvent property $R(\overline z)=R^*(z)$ and
the equality $Q(\overline z)=Q^*(z)$ which follows from
\eqref{q-fun}.
\end{rem}
First we prove some simple properties of the matrices $A$ and $B$:
\begin{lemma}
Let $A$, $B$ satisfy~\eqref{eq-ks1} and~\eqref{eq-ks2}, then
\begin{itemize}
\item[(a)] $\ker A^*\cap\ker B^*=0$, \item[(b)]
$\Lambda^{A,B}=\big\{(B^*x,A^*x),\  x\in V\}$.
\end{itemize}
\end{lemma}
\begin{proof}[{\bf Proof of Lemma}] (a) We first remark that the condition~\eqref{eq-ks2} is equivalent
to $\ran A+\ran B=V$. Then $\ker A^*\cap\ker B^*=\big(\ran
A)^\bot\cap\big(\ran B)^\bot= (\ran A+\ran B)^\bot=V^\bot=0$.

(b) Eq.~\eqref{eq-ks1} says that $\big\{(B^*x,A^*x),\ x\in
V\big\}\subset \Lambda^{A,B}$. At the same time, it follows
from~(a) that the linear subspace on the left-hand side has
dimension $n$, which coincides with the dimension of
$\Lambda^{A,B}$. Therefore, these two linear subspaces coincide.
\end{proof}

\begin{proof}[{\bf Proof of Proposition}]
(a) The matrices in question are adjoint to each other, therefore,
it suffices to prove that one of them is non-degenerate.

Consider first the case $z\in\CC\setminus\RR$. As follows from
\eqref{q-fun}, for such $z$ the matrix $\Im z\cdot\Im Q(z)$ is
positive definite (Here $\Im Q=(Q-Q^*)/(2i)$.) Assume that $\det\big(Q(z)B^*-A^*\big)=0$, then
there exists a non-zero $x\in V$ with
\begin{equation}
     \label{pr1}
(Q(z)B^*-A^*\big)x=0.
\end{equation}
If $B^*x=0$, then, due to item (a) of Lemma, we would have
$A^*x\ne0$, and \eqref{pr1} would be impossible. Therefore,
$B^*x\ne 0$. Taking the scalar product of $B^*x$ with both sides
of~\eqref{pr1} we get $\langle Q(z)B^*x,B^*x\rangle=\langle x,
AB^*x\rangle$. The number on the left-hand side has non-zero
imaginary part, while the number on the right-hand side is real
due to~\eqref{eq-ks2}. This contradiction proves the requested
non-degeneracy for non-real $z$.

Now let $z\in \RR\cap\res H^0$ and $\det\big(Q(z)B^*-A^*\big)=0$.
Let us show that $z\in\spec H^{A,B}$. Eq.~\eqref{q-fun} says that
$Q(z)$ is now self-adjoint. As the matrix $Q(z)B^*-A^*$ in
non-invertible, there is a non-zero $x\in
\ran\big(Q(z)B^*-A^*\big)^{\bot}=\ker\big(B Q(z)-A\big)$. By
definition, the element $\phi=\gamma_z x$ is an eigenvector of
$S^*$ corresponding to the eigenvalue $z$. Let us show that
$\phi\in\dom H^{A,B}$ (and then $z$ is an eigenvalue of
$H^{A,B}$). In fact, one has $\Gamma_1\phi=\Gamma_1\gamma_z
x=\Gamma_1 \Gamma_1^{-1}x=x$, $\Gamma_2\phi=\Gamma_2\gamma_z
x=Q(z)x$, and, therefore, $A\Gamma_1x-B\Gamma_2x=-\big(B
Q(z)-A\big)x=0$, which means that $\phi\in\dom H^{A,B}$.

(b) Taking into account the remark after Proposition it is enough to
prove only~\eqref{eq-krein1}. Actually, we only must show that for
$z\in\res H^0\cap\res H^{A,B}$ the linear relation inverse to $\gr
Q(z)-\Lambda^{A,B}$ is the graph of $B^*\big(Q(z)B^*-A^*\big)^{-1}$.
Taking into account item (b) of Lemma, we conclude that  $\dom
Q(z)\cap\dom \Lambda^{A,B}=\ran B^*$ and $\gr
Q(z)-\Lambda^{A,B}=\Big\{ \big(B^*u,(Q(z)B^*-A^*)u\big), \, u\in V
\Big\}=\Big\{ \big(B^*(Q(z)B^*-A^*)^{-1}w,w\big), \, w\in V\Big\}$.
\end{proof}
Let us consider some example from the point of view of the resolvent formula.
\begin{ex}[Graph with a single vertex]
Let $\HH=\oplus_{j=1}^n \HH_j$ with
$\HH_j=\LL^2(\RR_+^{(j)})$, where each $\RR_+^{(j)}$
is a copy of the positive half-line $[0,+\infty)$ . We will write the elements of
$\phi\in\HH$ in the vector form, $\phi(x)=\big(\phi_j(x_j)\big)$,
$\phi_j\in\HH_j$, $x_j\in\RR_+^{(j)}$.
By $S$ we denote an operator which acts on each $\HH_j$ as
$-d^2/dx_j^2$ with the domain $\dom S=\{ \phi=(\phi_j),\,
\phi_j\in W^{2,2}(\RR_+^{(j)}),\,\phi_j(0)=\phi'_j(0)=0,\,
j=1,\dots,n\}$. Then $S^*$ is the direct sum
$\oplus_{j=1}^n d^2/dx_j^2$ with the domain $\oplus_{j=1}^n W^{2,2}(\RR_+^{(j)}$.
An integration by parts show that as a boundary value space one can take
$(\CC^n,\Gamma_1,\Gamma_2)$ with
$\Gamma_1\phi=\Gamma_1(\phi_j)=\big(-\phi'_j(0)\big)\equiv
-\phi'(0)$,
$\Gamma_2\phi=\Gamma_2(\phi_j)=\big(\phi_j(0)\big)\equiv\phi(0)$;
details may be found in~\cite{ks,bg,exner}.
Then any self-adjoint extension of $S$ involves the boundary
conditions $A \phi'(0)+B\phi(0)=0$ with suitable $A$ and $B$,
which describes the coupling of $n$ half-lines at the origin. The
operator $H^0$ defined by $\phi'(0)=0$ is the direct sum of Neumann Laplacians, and its
Green function (the resolvent integral kernel) $G^0(x,y;z)$ is given by
\[
G^0(x,y;z)=\frac{1}{2\sqrt{-z}}\,\diag\big(
e^{-\sqrt{-z}\,|x_j-y_j|}+e^{-\sqrt{-z}\,(x_j+y_j)}\big),
\]
where the contiuous branch of the square root is chosen by the rule
$\Re \sqrt{z}>0$ for $z\notin (-\infty,0]$, and $x=(x_j)$,
$y=(y_j)$, $x_j,y_j\in \RR_+^{(j)}$. Then the elements
$g^j_z=G^0(\cdot,0;z)\, e_j=e^{-\sqrt{-z}\,x}/\sqrt{-z} \,e_j$,
$(e_j)$ is the standard basis in $\CC^n$, $j=1,\dots,n$,  form a basis in $\NN_z=\ker(S^*-z)$, and the map
$\gamma_z$ defined by $\CC^n\ni v=(v_j)\mapsto\big(v_j g^j_z \big)$ 
is the corresponding $\Gamma$-field, because $\Gamma_1\gamma_z v=v$
for any $v\in \CC^n$, and the $\mathcal{Q}$-function has the simple form
$Q(z)=1/\sqrt{-z}\, E_n$, where $E_n$ is the unit matrix of order $n$.
Therfore, the resolvent for $H^{A,B}$ takes the form
\[
R^{A,B}(z)=R^0(z)-\sum_{j,k=1}^n C_{jk}(z) \langle g^k_z,\cdot\rangle g^j_z
\]
with $C(z)=\big(BQ(z)-A\big)^{-1} B=\sqrt{-z}(B-\sqrt{-z}\,A)^{-1} B$.
A similar formula was obtained in~\cite{ks}.
\end{ex}

\begin{ex}[Point interactions with mixed boundary conditions]
Here we consider the well-known example of point interactions in three dimension~\cite{alb}.
Let $Y=(y_1,\dots,y_n)\subset \RR^3$. Denote by $S$ the Laplacian with the domain
$\dom S=\{\phi\in W^{2,2}(\RR^3),\, \phi(Y)=0\}$. This operator has deficiency indices $(n,n)$,
and the adjoint operator is the Laplacian with the domain
$\dom S^*=W^{2,2}(\RR^3\setminus Y)$. Each function $\phi\in \dom S^*$
has the following asymptotics:
\[
\phi(x)=\frac{1}{4\pi|x-y_j|}\,\phi^1_j+\phi^2_j +o(1), \quad x\to y_j,
\quad \phi^1_j,\phi^2_j\in\CC^n,\quad j=1,\dots,n,
\]
and the vectors $\Gamma_1 \phi= (\phi^1_j)$ and $\Gamma_2\phi=(\phi^2_j)$
can be considered as boundary values of $\phi$, see~\cite{pavlov,bg,kp}
for details. The operator $H^0$
is just the free Laplacian in $\RR^3$. Denote by $G^0(x,y;z)$ its Green function,
$G^0(x,y;z)=e^{-\sqrt{-z}|x-y|}/\big(4\pi|x-y|\big)$, then the functions
$g^j_z=G^0(\cdot,y_j;z)$, $j=1,\dots,n$, form a basis in $\ker (S^*-z)$, and the map
$\gamma_z:\,\CC^n\ni v=(v_j)\mapsto \sum v_j g^j_z$ is the $\Gamma$-field
(it is easy to check that
$\Gamma_1\gamma_z v= v$), and the $\mathcal{Q}$-function is given by the $n\times n$ matrix
\[
Q_{jk}(z)=G^0(y_j,y_k;z)=\frac{e^{-\sqrt{-z}\,|y_j-y_k|}}{4\pi|y_j-y_k|},\,j\ne k,
\quad Q_{jj}(z)=-\frac{\sqrt{-z}}{4\pi}.
\]
Therefore, the resolvent of the operator $H^{A,B}$ given by the boundary conditions
$A\phi^1=B\phi^2$ can be defined by its intergal kernel
\[
G^{A,B}(x,y;z)=G^0(x,y;z)-
\sum_{j,k=1}^n C_{jk}(z) G^0(x,y_j;z) G^0(y_k,y;z)
\]
with
$C(z)=\big(B Q(z)-A\big)^{-1}B$. We remark that the class of interactions described
by this formula is wider that the one studied in~\cite{alb}.
Some properties of $H^{A,B}$ in its dependence on $A$ and $B$
were studied recently in~\cite{kp}.
\end{ex}

\ack This work was partially supported by the
Sonderforschungsbereich~611 (Bonn), INTAS, and the Deutsche
Forschungsgemeinschaft.

\end{document}